\documentclass{article}

\PassOptionsToPackage{numbers}{natbib} % numeric citations to match \bibliographystyle{plain}
\usepackage[preprint]{neurips_2026}

\usepackage[utf8]{inputenc} % allow utf-8 input
\usepackage[T1]{fontenc}    % use 8-bit T1 fonts
\usepackage{hyperref}       % hyperlinks
\usepackage{url}            % simple URL typesetting
\usepackage{booktabs}       % professional-quality tables
\usepackage{amsfonts}       % blackboard math symbols
\usepackage{amsmath}
\usepackage{amssymb}        % \bigstar (Figure 8 caption)
\usepackage{nicefrac}       % compact symbols for 1/2, etc.
\usepackage{microtype}      % microtypography
\usepackage{xcolor}         % colors
\usepackage{graphicx}
\usepackage{tabularx}
\usepackage{array}
\usepackage{textcomp}
\usepackage{xspace}
\usepackage{listings}
\usepackage{enumitem}
\usepackage[section]{placeins} % keep floats inside their section
\usepackage{needspace}         % keep short listings on one page
\usepackage{float}             % [H]: Figure 1 stays on the first page

\graphicspath{{../figs/}}

\hypersetup{colorlinks=true, linkcolor=blue!60!black, citecolor=blue!60!black, urlcolor=blue!60!black}

\lstdefinestyle{py}{morecomment=[l]{\#}}

\newcommand{\us}{\,\textmu{}s\xspace}
\newcommand{\code}[1]{\texttt{#1}}
\newcolumntype{L}{>{\raggedright\arraybackslash}X}
\newcolumntype{P}[1]{>{\raggedright\arraybackslash}p{#1}}

\title{Exploiting Native Hardware Heterogeneity: Custom GPSIMD Operators for Accelerating Emerging LLM Memory on AWS Trainium}

\author{%
  Zifan He, Jason Cong \\
  University of California, Los Angeles \\
  \texttt{zifanhe1202@g.ucla.edu, cong@cs.ucla.edu} \\
}

\begin{document}

\maketitle

\begin{abstract}
Emerging memory modules of large language models, such as Engram in DeepSeek-V4.1-Flash, mix irregular integer work (a 64-bit n-gram hash and a sparse table lookup) with dense computation (a GEMM and a gate). When deploying on a GPU-FPGA system, we can disaggregate operations to different devices for Engram to speedup by 5.9$\times$. AWS Trainium provides a similar heterogeneity natively: each Trainium  NeuronCore pairs dense engines with an customizable GPSIMD Engine, and all engines share the on-chip SBUF. However, the Neuron Kernel Interface (NKI) exposes the GPSIMD cores only through fixed built-in instructions, and the \textbf{official GPSIMD custom-operator toolchain does not support Trainium 2 or 3 and cannot be used with NKI}. In this work, we build a toolchain\footnote{\url{https://github.com/OswaldHe/neuron-gpsimd-tool}} that compiles C++ custom operators for the GPSIMD cores and injects them into the compiled NEFF from NKI kernel on Trainium 3, expanding the customizability of GPSIMD with current software stack. Over a hand-optimized NKI baseline of DeepSeek-V4.1-Flash, two GPSIMD operators speed up the n-gram hashing by 1.79$\times$ at decode, the table retrieval by up to 4.95$\times$, and the whole Engram computation by up to 1.14$\times$.
\end{abstract}

\begin{center}
\fbox{\parbox{0.95\linewidth}{\small\textbf{Disclaimer:} Every Trainium conclusion and experiment in this work comes from the Neuron SDK 2.33 toolchain, run and measured on a single trn3.48xlarge instance. Everything is from open sourced code and packages. We installed the Neuron GPSIMD Toolchain v1.25, which is a legacy version used for trn1 instances. We then built the toolchain based on how the Neuron runtime loads and runs GPSIMD code. Therefore, all of the information should be treated as reference only. The information does not necessarily reflect how AWS implements things internally.}}
\end{center}

\begin{figure}[H]
  \centering
  \includegraphics[width=0.55\linewidth]{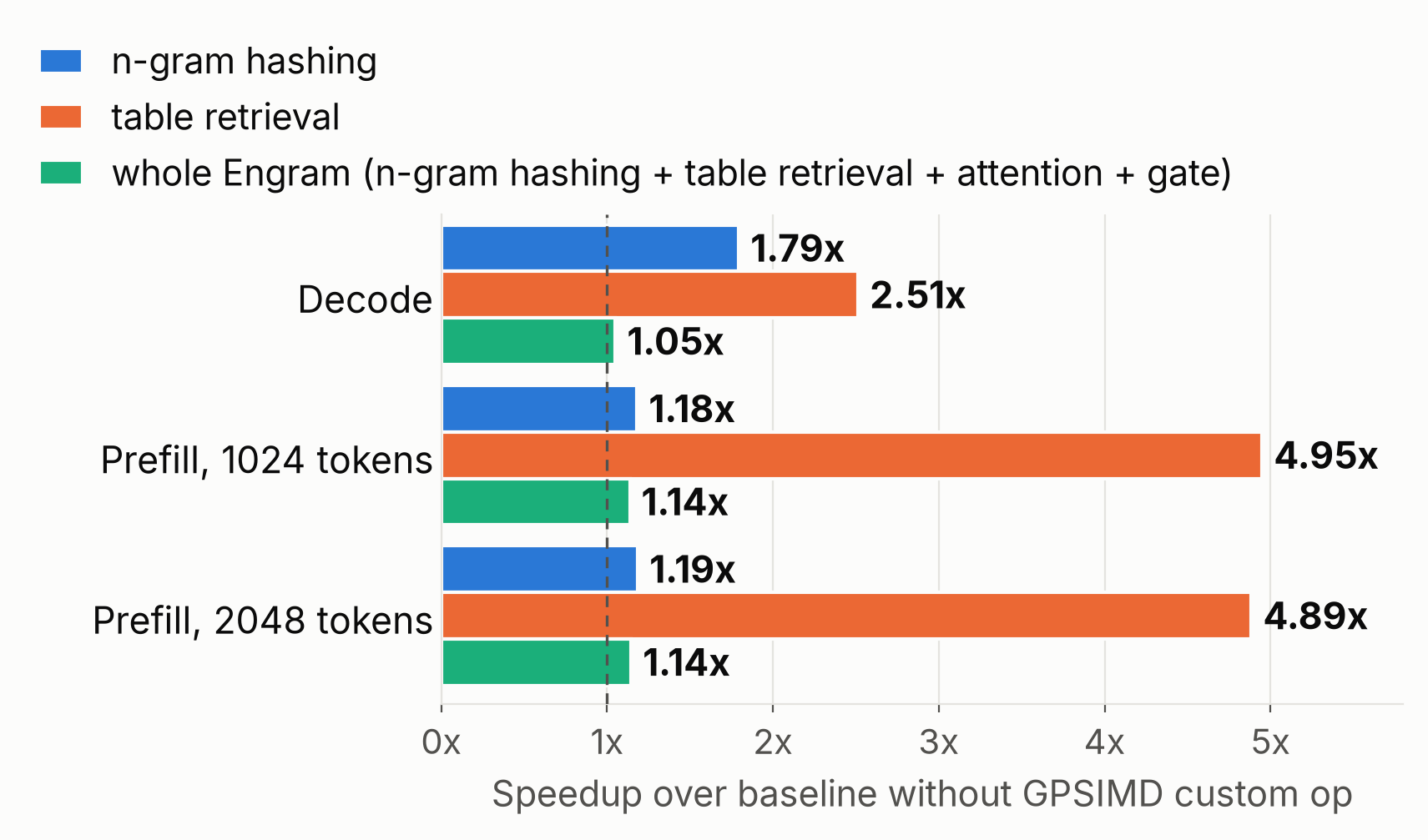}
  \caption{Speedup over the hand-optimized NKI baseline of vLLM Neuron on one NeuronCore.}
  \label{fig:speedups}
\end{figure}
\newpage

{\small\tableofcontents}
\newpage

%=====================================================================
\section{Introduction}
\label{sec:intro}

\subsection{Heterogeneity of Computation Properties in LLMs}
\label{sec:hetero}

\begin{figure}[!htbp]
  \centering
  \includegraphics[width=0.7\linewidth]{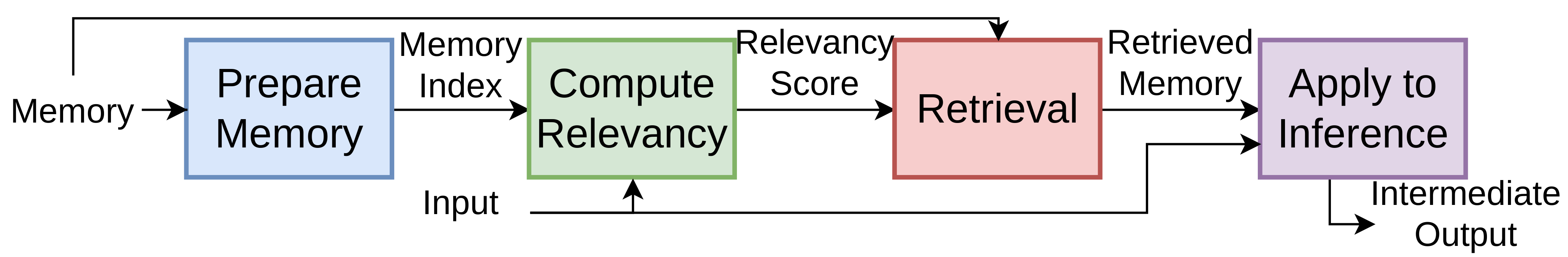}
  \caption{The four-step memory processing pipeline in LLMs, from~\citep{he2026memory}.}
  \label{fig:pipeline}
\end{figure}

Modern LLMs depend on complex mechanisms to manage their memory: sparse attention selects key-value entries, retrieval-augmented generation (RAG) selects documents, and compressed memories select memory slots. Prior works published in ICML 2026~\citep{he2026memory} showed that these mechanisms share a \textbf{memory processing pipeline} of four steps (Figure~\ref{fig:pipeline}):

\begin{enumerate}[leftmargin=*,itemsep=1pt]
  \item \textbf{Prepare Memory}: turn the memory into a form that is easy to search and use (for example, indexing vectors).
  \item \textbf{Compute Relevancy}: score or identify the memory entries that matter for the current input.
  \item \textbf{Retrieval}: select and fetch the relevant subset.
  \item \textbf{Apply to Inference}: feed the retrieved memory into the rest of the model.
\end{enumerate}

The key is that these steps have very different computational characteristics. Some steps are dense and regular computations, whereas others are irregular and data dependent. Therefore, a natural approach of acceleration is \textbf{fine-grain disaggregated} inference on \textbf{heterogeneous platforms}. He et al.~\citep{he2026memory} apply a GPU-FPGA heterogeneous system~\citep{heterollm}: the dense steps stayed on the GPU and the irregular steps moved to the FPGA, for up to 2.2$\times$ faster inference and up to 4.7$\times$ lower energy. Such methodology is also transferable to more recent LLMs.

\subsection{Example: Engram in DeepSeek-V4.1-Flash, and How We Accelerate It on a GPU-FPGA System}
\label{sec:engram}

DeepSeek-V4.1-Flash~\citep{deepseek_v41_flash} is a 40-layer mixture-of-experts (MoE) reasoning model from DeepSeek. It combines sparse attention with MoE layers of 384 routed experts, 6 of which serve each token. Layers 1 and 14 also have an \textbf{Engram} module~\citep{engram} that gives the model a large n-gram memory. For each token, Engram hashes the token and its preceding tokens into row ids, reads these rows from a large learned table, and adds a gated projection of the rows to the hidden state. Engram maps to the memory processing pipeline as follows:

\begin{itemize}[leftmargin=*,itemsep=1pt]
  \item \textbf{Prepare Memory.} The table is trained offline. In DeepSeek-V4.1-Flash, each Engram layer has 384M rows of 256 bf16 values. At run time, each token's n-gram window is collected: the token's compressed id and the ids of the 3 preceding tokens.
  \item \textbf{Compute Relevancy.} An n-gram hash turns each window into row ids. For $n = 2, 3, 4$, the hash multiplies each token id by a 64-bit constant, combines the products with XOR, and reduces the result modulo one of 8 primes of about 16M. Each token gets $3 \times 8 = 24$ row ids per layer.
  \item \textbf{Retrieval.} The 24 rows of each token are read into an embedding \code{emb} of shape $[T, 24 \times 256]$.
  \item \textbf{Apply to Inference.} A GEMM projects \code{emb} to 4 keys and 1 value to compute the attention score between the hidden state and the keys. A gate adds the value to each of the 4 residual streams.
\end{itemize}

We also tried to deploy Engram on our GPU-FPGA heterogeneous system to demonstrate the effectiveness of such abstraction~\citep{heterollm}. Here we use the original demo of Engram in~\citep{engram_code}:

\begin{itemize}[leftmargin=*,itemsep=1pt]
  \item \textbf{FPGA: Compute Relevancy, Retrieval, and the key and value projections.} A hash unit computes the 16 row ids of one token per cycle. 16 lookup units read the rows from separate HBM channels. 8 projection units keep the weights in on-chip memory.
  \item \textbf{GPU: the gate, a short convolution, and the residual add.} These steps need the hidden state from the previous layer.
\end{itemize}

\begin{figure}[!htbp]
  \centering
  \includegraphics[width=0.85\linewidth]{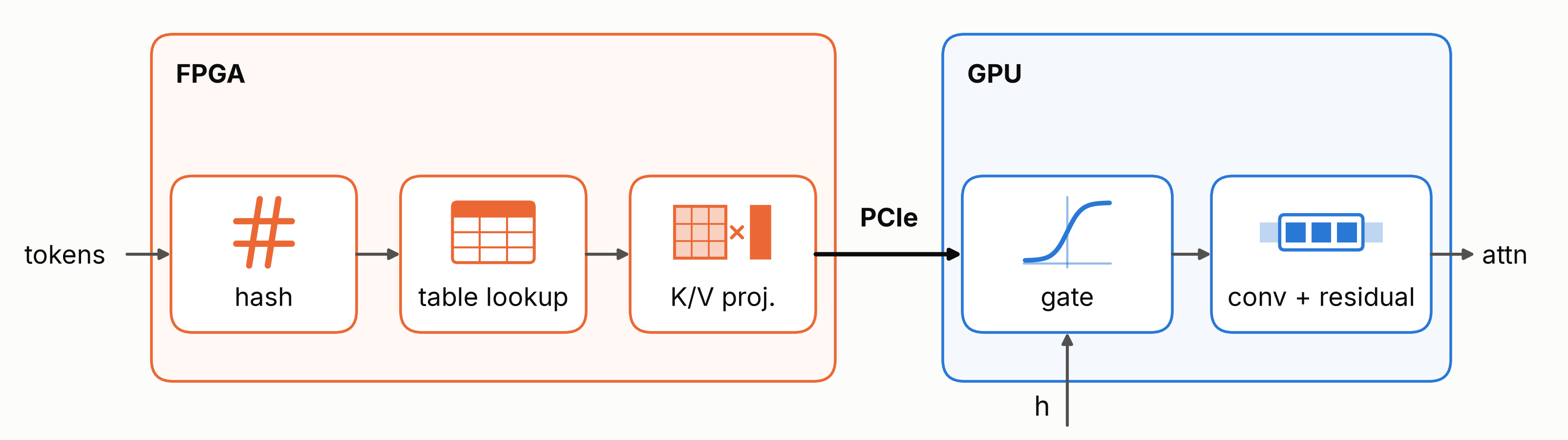}
  \caption{The GPU-FPGA split of Engram in HeteroLLM~\citep{heterollm}. $h$ is the hidden state.}
  \label{fig:fpga_split}
\end{figure}

\begin{figure}[!htbp]
  \centering
  \includegraphics[width=0.55\linewidth]{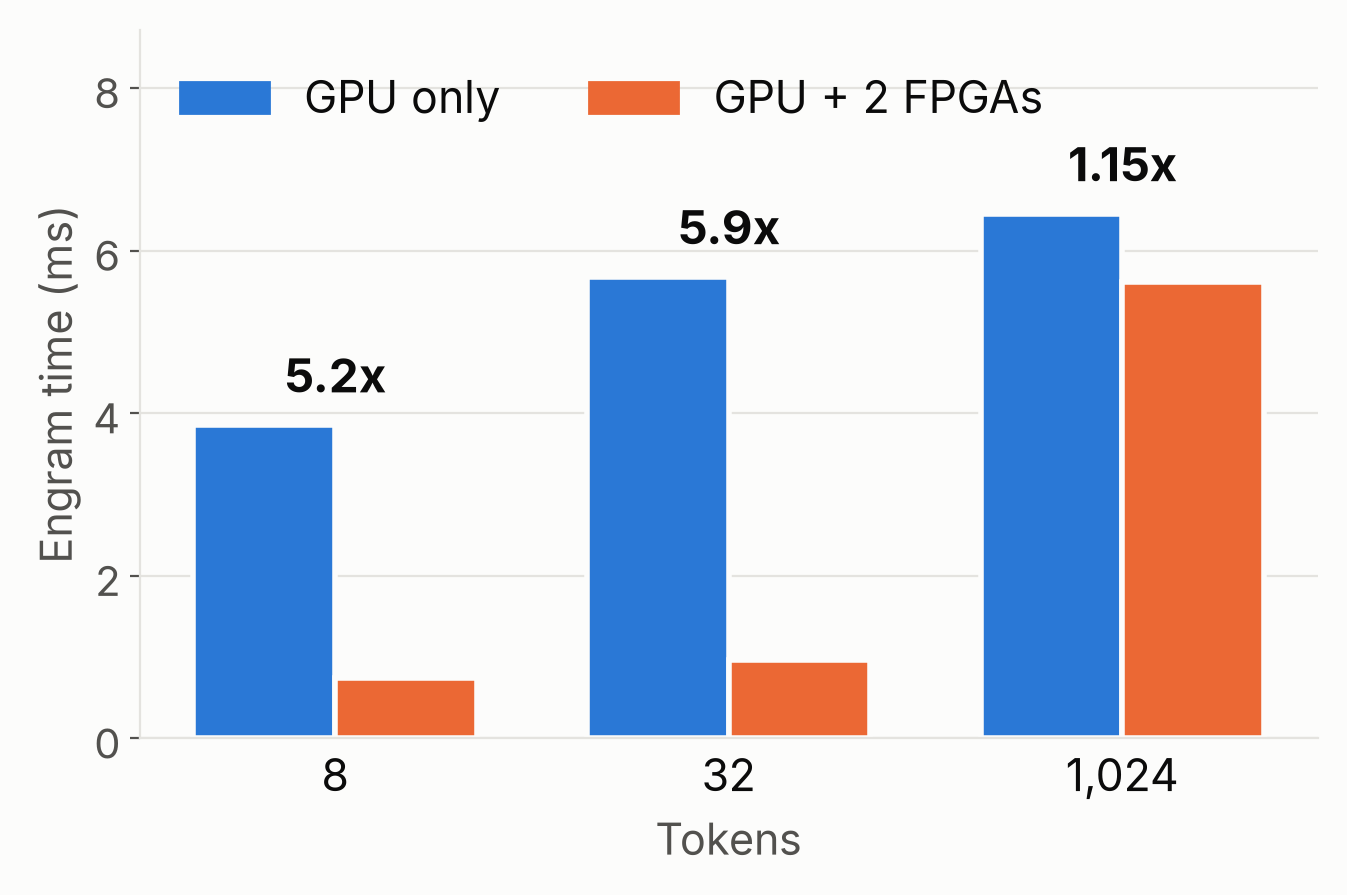}
  \caption{Engram on the GPU alone and on the GPU with 2 FPGA boards. Engram reference configuration~\citep{engram_code}.}
  \label{fig:fpga_results}
\end{figure}

The baseline is the reference code~\citep{engram_code} on the GPU, which computes the hash on the host CPU. With 2 AMD U55C FPGAs and 1 AMD MI210 GPU, Engram is 5.2$\times$ faster at 8 tokens, 5.9$\times$ faster at 32 tokens and 1.15$\times$ faster at 1024 tokens (Figure~\ref{fig:fpga_results}). The gain is largest with short prefill and decode.

The prototype shows that Engram gains from moving its irregular steps to flexible hardware. This technique is transferable to AWS Trainium devices: Trainium NeuronCore is natively heterogeneous.

\subsection{Native Hardware Heterogeneity of Trainium}
\label{sec:trainium}

\begin{figure}[!htbp]
  \centering
  \includegraphics[width=0.85\linewidth]{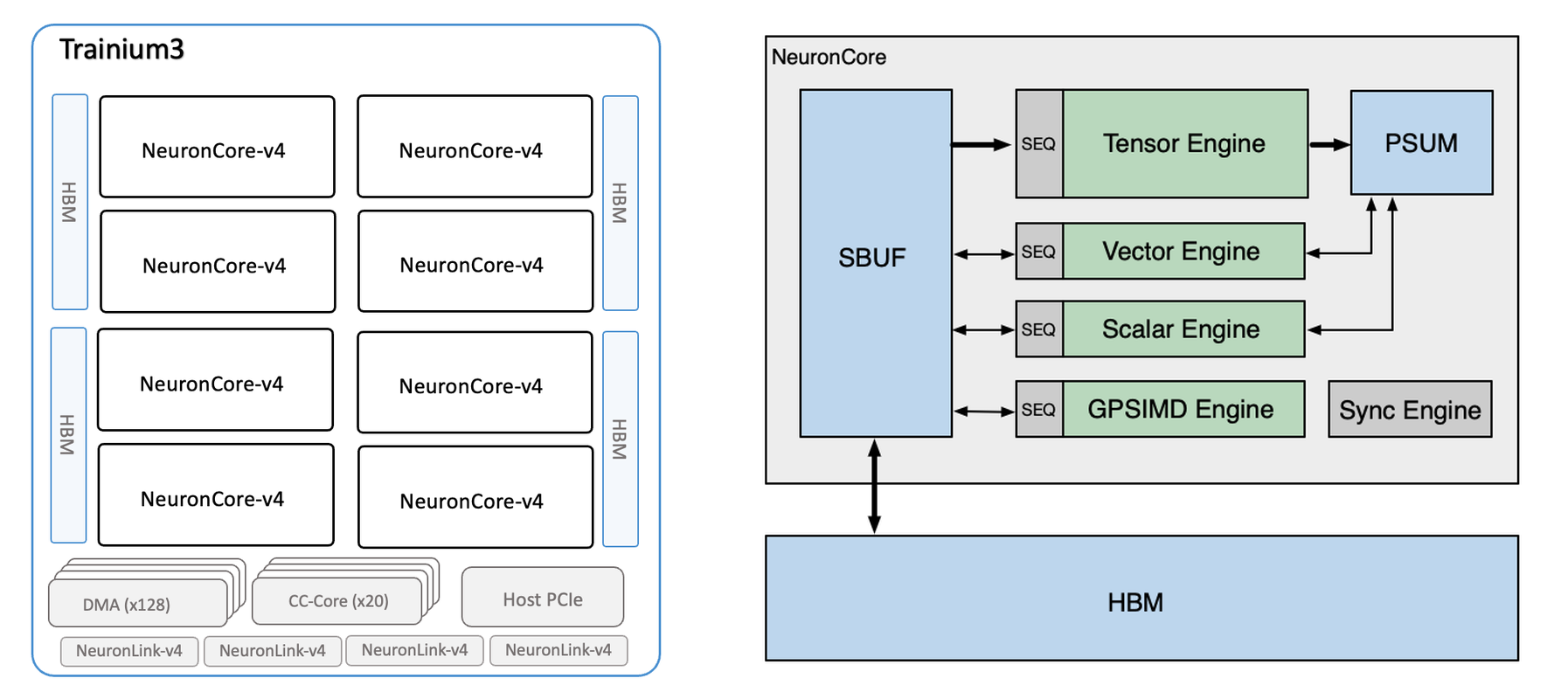}
  \caption{Trainium~3 device diagram and the architecture of a NeuronCore, from the Neuron SDK documentation~\citep{neuron_trn3_arch,neuron_trn3_nki_arch}.}
  \label{fig:trn3}
\end{figure}

A trn3.48xlarge instance exposes 64 NeuronCores (each made of 2 physical cores). A Trainium~3 NeuronCore is a heterogeneous processor (Figure~\ref{fig:trn3}). The \textbf{Tensor Engine} is a 128$\times$128 systolic array for matrix multiplication. The \textbf{Vector Engine} and the \textbf{Scalar Engine} apply element-wise operations, reductions and activation functions to tiles of up to 128 partitions. The \textbf{Sync Engine} issues queued DMA transfers. All engines read and write the on-chip \textbf{SBUF} (state buffer), and the Tensor Engine accumulates into \textbf{PSUM}.

The fifth engine, the \textbf{GPSIMD Engine}, is different from the other four. The GPSIMD Engine is a group of 8 general-purpose SIMD cores. Each core has its own local data RAM, an integrated DMA channel (iDMA), and a compute unit that support 64-bit wide arithmetics. The cores run firmware that implements the GPSIMD Engine's instructions in software.

Each engine fits a different kind of computation. Hence, each step of Engram has a natural place (Table~\ref{tab:engines}).

\begin{table}[h]
  \centering
  \caption{The engines of a Trainium~3 NeuronCore and the Engram computations that suit them.}
  \label{tab:engines}
  \small
  \begin{tabularx}{\linewidth}{@{}P{2.9cm}P{4.2cm}L@{}}
    \toprule
    \textbf{Engine} & \textbf{Specialized computations} & \textbf{Which computation in Engram?} \\
    \midrule
    Tensor Engine & dense matrix multiplication & the dense GEMM that projects the retrieved embeddings to the keys and the value \\
    Vector and Scalar Engines & element-wise floating-point math, reductions, activations & the gating network \\
    DMA/Sync engines & bulk copies between HBM and SBUF & reading the table rows \\
    GPSIMD Engine & integer math, per-element branches, variable-length output & the n-gram hashing, and finding which lookups of the table retrieval hit this rank's table shard \\
    \bottomrule
  \end{tabularx}
\end{table}

The n-gram hashing and the table retrieval go to the GPSIMD cores because both are irregular integer work. The hash needs 64-bit products and a 64-bit modulo: the cores compute the products, while the other engines compute exact integer arithmetic only on 32-bit values. In the table retrieval, only about 1.5\% of a token's lookups hit the table shard of a rank. GPSIMD engine can drive the table lookups on these entries in the table to issue DMA. The dense GEMM and the gate stay on the Tensor, Vector and Scalar Engines, as GPSIMD engine has much lower peak performance in dense linear algebra than other engines. In general, a step suits the GPSIMD cores when it is integer-heavy, data-dependent, and small in data compared with the dense work around it.

Although this is a promising plan, a major blocker for developers on Trainium is that there is no official support to connect custom GPSIMD operations with the kernel language in Trainium to optimize the whole kernel. In the next section, we will illustrate how to integrate custom GPSIMD operator with Neuron Kernel Interface (NKI) for a coherent optimization.

%=====================================================================
\section{Build and Use GPSIMD Custom Operators for Kernel Optimization}
\label{sec:toolchain}

The \textbf{Neuron Kernel Interface (NKI)}~\citep{nki_about} is the Python kernel language for Trainium to speedup computations. Currently, NKI exposes the GPSIMD Engine only through built-in instructions on Trainium~3: tensor operations with \code{engine=gpsimd}, \code{iota}, \code{topk}, and the descriptor generation of indirect DMAs. The official toolchain for custom GPSIMD code, the Neuron custom C++ operators~\citep{neuron_customops}, does not solve this problem on Trainium 2 and Trainium 3. First, it supports only \textbf{NeuronCore-v2} (Trainium~1 and Inferentia~2), and the compiler rejects it for Trainium~3 targets (error \code{NCC\_EVRF030}). Second, \textbf{a custom operator runs only as a PyTorch operator through PyTorch/XLA, and it cannot be called from an NKI kernel}. Third, the toolchain is in maintenance mode starting with Neuron 2.33. Therefore, Trainium~2 and Trainium~3 have no official way to add a custom GPSIMD operator to a kernel. To fill this gap, we reverse engineered how the Neuron runtime loads and runs GPSIMD code, and built a toolchain that injects GPSIMD operators into the Neuron Executable File Format (NEFF) of a compiled NKI kernel.

\subsection{Overall Development and Deployment Flow}
\label{sec:flow}

\begin{figure}[!htbp]
  \centering
  \includegraphics[width=0.8\linewidth]{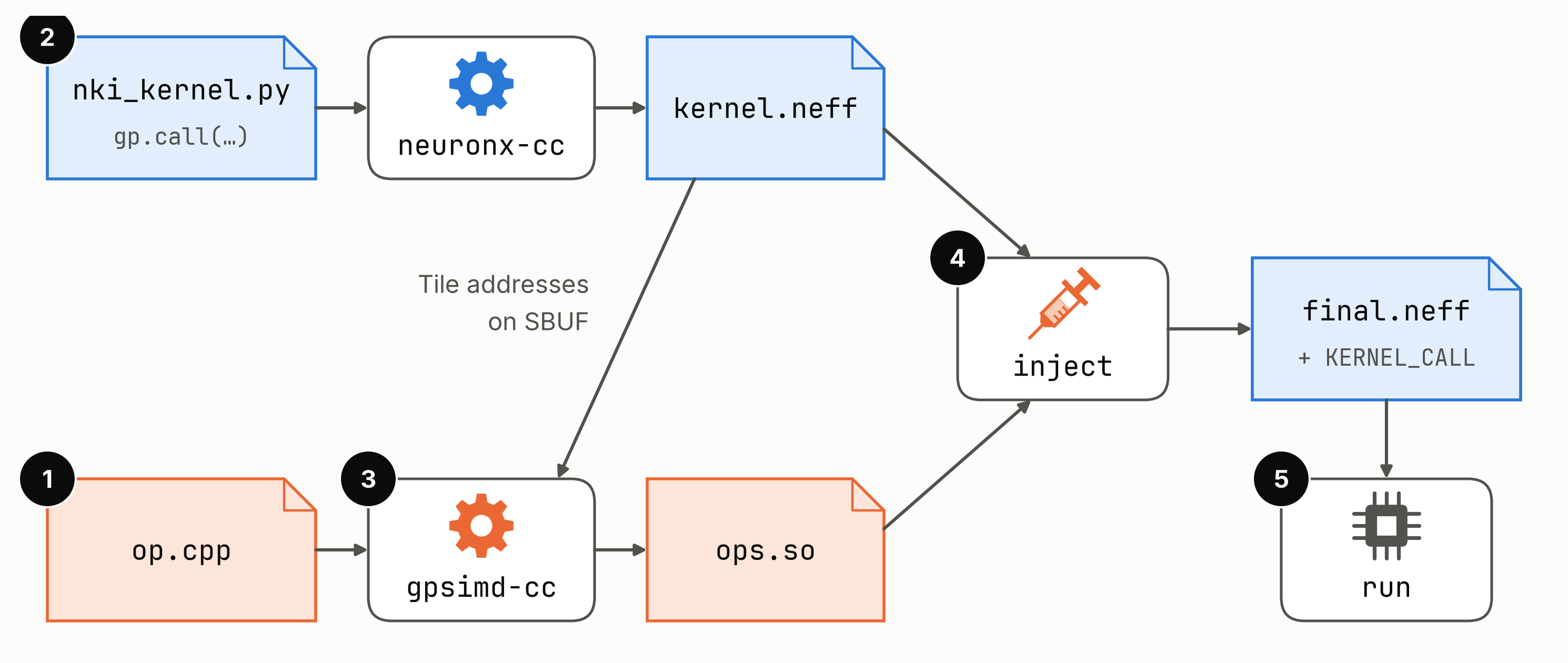}
  \caption{Overall development and deployment flow of building and using GPSIMD custom operator and connect with optimized NKI kernel.}
  \label{fig:toolchain}
\end{figure}

The Neuron compiler turns an NKI kernel into a NEFF, which holds one instruction stream per engine~\citep{neff_doc}. Our toolchain does not change the compiler. The NKI kernel is compiled as usual, and the toolchain then edits the compiled NEFF to place each custom operator in the GPSIMD Engine's instruction stream. The rest of the NEFF, including the compiler's schedule of the other engines, stays as the compiler produced it. Figure~\ref{fig:toolchain} shows the flow in five steps:

\begin{enumerate}[leftmargin=*,itemsep=1pt]
  \item \textbf{Write} the operator in C++ using the \code{gp::} namespace with the \code{gpsimd/gpsimd.hpp} header.
  \item \textbf{Mark} the call site in the NKI kernel with \code{gp.call(...)}, and compile the kernel with the NKI compiler as usual.
  \item \textbf{Build} the operator library: the toolchain reads the SBUF location of each tile from the compiled NEFF and compiles the C++ operator for the GPSIMD cores so that it can access the corresponding SBUF address.
  \item \textbf{Inject} the library into the NEFF, and replace the marked instructions with one \code{KERNEL\_CALL} instruction.
  \item \textbf{Load and run} the NEFF with the Neuron runtime, which will load the shared object for the operator when hit the \code{KERNEL\_CALL} and run on GPSIMD.
\end{enumerate}

\code{gpsimd-build} in our toolchain will handle all the compilation and injection process to produce the final NEFF file. The user only need to provide the C++ file for the GPSIMD operator and the NKI kernel python file with \code{gp.call()} that call the function in GPSIMD operator.

\subsection{Write Operators in C++}
\label{sec:write}

A GPSIMD operator is a C++ function that can express the execution of all 8 cores (SPMD similar to CUDA kernel or NKI SPMD for LNCs). The runtime library gives each core typed views of the SBUF tiles passed by \code{gp.call}, its share of the partitions, and asynchronous iDMA copies between SBUF, HBM and the core's local buffer. The example below adds two int32 tiles:

\needspace{15\baselineskip}
\begin{lstlisting}[language=C++]
#include <gpsimd/gpsimd.hpp>
GPSIMD_OP(vadd_i32) {                         // extern "C" void vadd_i32(gp_call *), runs on all 8 cores
    gp::Call c(call);
    auto a = c.in<int32_t>(0), b = c.in<int32_t>(1);      // SBUF tiles passed by gp.call
    auto o = c.out<int32_t>(0);
    const uint32_t n = o.elems();
    int32_t la[4096], lb[4096];                           // staging buffers in local data RAM
    for (gp::Range blk : c.blocks(o.partitions(), 4096 / n)) {   // this core's partitions
        c.fetch(a, blk.lo, blk.size(), la);                       // iDMA SBUF -> local RAM
        c.dma_wait(c.fetch(b, blk.lo, blk.size(), lb));
        for (uint32_t k = 0; k < blk.size() * n; ++k) la[k] += lb[k];
        c.dma_wait(c.flush(o, blk.lo, blk.size(), la));           // iDMA local RAM -> SBUF
    }
}
\end{lstlisting}

The operator can access the core ID with \code{c.core()} (0 to 7) and the number of cores with \code{c.ncores()} (8). Most operators use the index to access partition of data through \code{c.split(n)}, which gives core $k$ the $k$-th of 8 equal blocks of $[0, n)$, or \code{c.blocks(n, rows)}, which walks that block in chunks of at most \code{rows} partitions. In the example above, each core takes 16 of the 128 partitions. The same split written by hand, with direct element access to SBUF instead of iDMA staging:

\begin{lstlisting}[language=C++]
const uint32_t k = c.core(), K = c.ncores();             // this core (0..7) and the number of cores (8)
const uint32_t P = o.partitions(), per = (P + K - 1) / K; // 16 partitions per core for P = 128
for (uint32_t p = k * per; p < (k + 1) * per && p < P; ++p)
    for (uint32_t i = 0; i < n; ++i)
        o.store(p, i, a.load(p, i) + b.load(p, i));       // each SBUF load takes about 296 cycles
if (k == 0) gp_printf("done: %u partitions\n", P);        // work for one core only
\end{lstlisting}

In practice, an operator rarely accesses SBUF element by element. As in the first example, it copies a block of its partitions into local RAM with \code{c.fetch}, computes there, and writes the result back with \code{c.flush}, because local RAM is much faster to access than SBUF. Table~\ref{tab:api} lists the functions of the runtime library.

\begin{table}[!htbp]
  \centering
  \caption{Functions of the operator runtime library.}
  \label{tab:api}
  \footnotesize
  \begin{tabularx}{\linewidth}{@{}P{2.1cm}LP{4.0cm}@{}}
    \toprule
    \textbf{Group} & \textbf{Functions} & \textbf{Purpose} \\
    \midrule
    Cores and work split & \code{c.core()}, \code{c.ncores()}, \code{c.split(n)}, \code{c.partitions(t)}, \code{c.blocks(n, rows)} & this core's index and its share of the work \\
    Tiles & \code{c.in<T>(k)}, \code{c.out<T>(k)} & views of the SBUF tiles passed by \code{gp.call}; \code{T} is \code{int8\_t} to \code{uint32\_t}, \code{gp::f32}, \code{gp::bf16} or \code{gp::f16} \\
    Element access & \code{t.load(p, i)}, \code{t.store(p, i, v)}, \code{t.row(p)}, \code{t.partitions()}, \code{t.elems()}, \code{t.contiguous()}, \code{t.span()}, \code{t.offset(i)} & element \code{i} of partition \code{p}, row-major over the free dimensions \\
    iDMA copies & \code{c.fetch(t, p, n, local)}, \code{c.flush(t, p, n, local)}, \code{c.dma(dst, src, bytes)}, \code{c.dma\_2d(dst, src, row\_bytes, nrows, src\_pitch, dst\_pitch)}, \code{c.dma\_wait(ticket)}, \code{c.dma\_wait\_all()} & asynchronous copies between local RAM, SBUF and HBM, completed in order \\
    HBM & \code{gp::hbm\_ptr<T>(params, k)}, \code{gp::hbm\_mapped(params)} & direct access to HBM tensors that the runner maps \\
    Cooperation across cores & \code{c.barrier()}, \code{c.allreduce\_add(v)}, \code{c.allreduce\_max(v)}, \code{c.allreduce\_min(v)}, \code{c.exscan\_add(v, \&total)}, \code{c.lock()}, \code{c.unlock()}, \code{c.atomic\_add(word, v)}, \code{c.shared(\&n)} & synchronization and data exchange through the call's SBUF scratch tile; all cores must call them in the same order \\
    Integer & C++ integer arithmetic, including 64-bit \code{* / \%}; \code{memcpy}, \code{memmove}, \code{memset}, \code{memcmp}; inline assembly of base-ISA instructions such as \code{muluh} & 64-bit operations run as runtime helpers \\
    Floating point & \code{gp::f32} (\code{+ - * /}, comparisons, \code{to\_i32()}), \code{gp::fma}, \code{sqrt}, \code{min}, \code{max}, \code{abs}, \code{floor}, \code{ceil}, \code{trunc}, \code{rint}, \code{round}, \code{ldexp}; \code{gp::bf16}, \code{gp::f16}; \code{GP\_F32(x)} & IEEE float32 in software, bit-identical to numpy \\
    Math & \code{gp::math::exp}, \code{log}, \code{tanh}, \code{sigmoid}, \code{sin}, \code{cos}, \code{gelu}, \code{rsqrt}, \code{isnan}, \code{isinf}, \code{copysign}, \code{frexp} & float32 functions \\
    \bottomrule
  \end{tabularx}
\end{table}

Next, we show how an NKI kernel calls this function.

\subsection{Integrate the Custom Operator with NKI}
\label{sec:integrate}

In the NKI kernel, \code{gp.call} takes the operator name and the SBUF tiles that the operator reads and writes:

\needspace{11\baselineskip}
\begin{lstlisting}[language=Python,style=py]
import gpsimd_nki as gp

@nki.jit
def kernel(a, b):
    out = nl.ndarray((128, 512), dtype=nl.int32, buffer=nl.shared_hbm)
    ta, tb, tc = (nl.ndarray((128, 512), dtype=nl.int32, buffer=nl.sbuf) for _ in range(3))
    nisa.dma_copy(dst=ta, src=a); nisa.dma_copy(dst=tb, src=b)
    gp.call("vadd_i32", ins=[ta, tb], outs=[tc])        # runs on the 8 GPSIMD cores
    nisa.dma_copy(dst=out, src=tc)
    return out
\end{lstlisting}

Because the NKI compiler does not know custom operators, \code{gp.call} cannot emit the call itself. Instead, it emits \textbf{marker instructions}: ordinary GPSIMD Engine instructions (\code{TENSOR\_SCALAR\_ARITH\_OP} and \code{MEMSET}) whose immediate operand carries a tag (an encoded bit value) so that the injector will not confuse with other instructions. The tag tells the toolchain which call site, role and tile a marker belongs to. The marker indicates what the GPSIMD operator will access and modify (Table~\ref{tab:markers}).

\begin{table}[h]
  \centering
  \caption{Marker instructions emitted by \code{gp.call}.}
  \label{tab:markers}
  \small
  \begin{tabularx}{\linewidth}{@{}lP{4.6cm}L@{}}
    \toprule
    \textbf{Marker} & \textbf{What it does} & \textbf{What the compiler then does for the operator} \\
    \midrule
    touch & updates each input tile in place & allocates the input tiles in SBUF and keeps them alive until the call \\
    join & fully writes each output tile & allocates the output tiles \\
    chain & links the markers through a small token tile & orders every producer of an input before the call and every consumer of an output after it \\
    fill & writes a scratch tile & reserves SBUF scratch for cooperation across cores \\
    \bottomrule
  \end{tabularx}
\end{table}

Next, the toolchain analyzes the compiled NEFF and injects the operator into it. A NEFF is a container with a 1024-byte header followed by a gzip-compressed tar archive:

\begin{lstlisting}
kernel.neff
|-- [header, 1024 B]          payload size, MD5 of the payload, and other metadata
`-- [payload, tar.gz]
    |-- neff.json             top-level graph: input tensors -> kernel -> output tensors
    |-- kelf-0.json           list of subgraphs and the target ("mariana" = Trainium 3)
    |-- info.json             compiler and NKI versions
    `-- sg00/                 program of one physical core
        |-- def.json          memory variables, DMA queues and their semaphores, engine-to-stream map
        |-- tensor_map.json   data type, shape and layout of each input and output tensor
        |-- PE0.bin           Tensor Engine instruction stream
        |-- Activation0.bin   Scalar Engine instruction stream
        |-- DVE0.bin          Vector Engine instruction stream
        |-- Pool0.bin         GPSIMD Engine instruction stream
        |-- SP0.bin           Sync Engine instruction stream
        |-- {PE0,Activation0,DVE0,Pool0,SP0}.json             description of each stream
        |-- act_info.json, <set>.json, <set>_{bkt,ctrl}.bin   activation tables of the Scalar Engine
        |-- default_*_table.bin                               Vector Engine microcode tables
        `-- debug_info_*.dbg                                  debug information
\end{lstlisting}

\begin{figure}[!htbp]
  \centering
  \includegraphics[width=\linewidth]{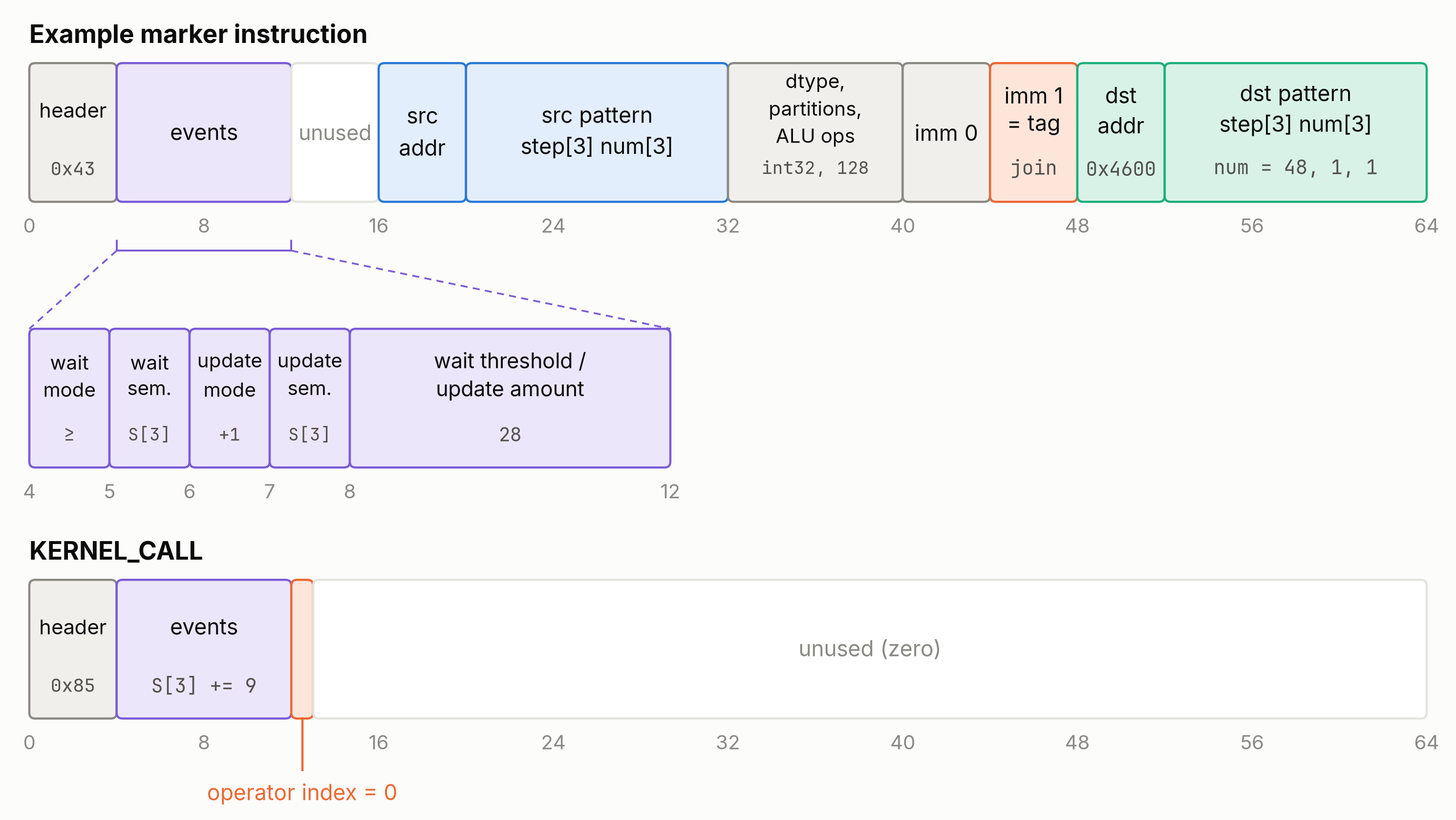}
  \caption{Layout of the 64-byte instructions that the toolchain reads and writes, with the values of the n-gram hashing call site. Top: the join marker emitted by \code{gp.call}, with its events block enlarged. Bottom: the \code{KERNEL\_CALL} that replaces the markers.}
  \label{fig:slot}
\end{figure}

\code{neff.json} describes the NEFF as a graph from the input tensors through the kernel to the output tensors, \code{kelf-0.json} lists the subgraphs, and \code{info.json} records the compiler versions. Each instruction stream (\code{<engine>0.bin}) is an array of 64-byte instructions. Figure~\ref{fig:slot} shows the layout of the two instructions that the toolchain reads and writes, with the values of the n-gram hashing kernel. The first 12 bytes have the same layout for every opcode: a header that starts with the opcode, and the \textbf{events block}, which holds one semaphore wait and one semaphore update. The rest of the slot holds the operands. A tensor operand is an address or an access pattern in NKI's \code{.ap()}.

The engines run independently and synchronize only through \textbf{semaphores}. When an instruction finishes, it can add to a semaphore: GPSIMD Engine instructions use S[3], and DMA transfers use the semaphores of their queue. A wait such as \code{S[3] >= 28} in Figure~\ref{fig:slot} counts the increments since the start of the kernel. Thus, the schedule of the whole kernel is encoded in these counts. Adding or removing an instruction that increments a semaphore shifts every later wait on that semaphore. Consequently, injection must modify the events bytes in the instructions.

\code{gpsimd-neff analyze} reads the compiled \code{kernel.neff} and finds the tagged markers in the GPSIMD Engine's instruction stream (\code{Pool0.bin}). The marker operands give the SBUF location of every tile: the first partition, the number of partitions, the byte offset inside a partition, the data type, and the free-axis access pattern. In the join marker of Figure~\ref{fig:slot}, which writes the output tile of the n-gram hashing call site, the second immediate holds the tag $2^{20} + 512\cdot 5$ (site 0, role 5: join). The destination address is \code{0x4600}, and the operand covers 128 partitions with 48 int32 values each. Correspondingly, we hardcode the SBUF addresses in the custom operator.

\code{gpsimd-neff inject} adds two files and one key to the subgraph folder, rewrites the instruction streams, and repacks the archive:

\begin{lstlisting}
final.neff
`-- sg00/
    |-- Pool0.json               + "gpsimd_kernels_def": "gpsimd_kernels_def.json"
    |-- gpsimd_kernels_def.json  + [{"name": "ngram_hash__s0", "lib": "/sg00/ops.so"}]
    |-- ops.so                   + the operator library (Xtensa ELF)
    `-- Pool0.bin                   markers replaced by one KERNEL_CALL per call site
\end{lstlisting}

In \code{Pool0.bin}, the markers of each call site become one \code{KERNEL\_CALL} (opcode \code{0x85}) at the position of the site's first output marker. Byte 12 of the \code{KERNEL\_CALL} is the index of the operator in \code{gpsimd\_kernels\_def.json}, and the other operand bytes are zero. The other markers become \code{EVENT\_SEMAPHORE} instructions that keep the markers' waits, and the fill marker becomes a real \code{MEMSET} that clears the scratch tile. The bytes below come from the n-gram hashing kernel of the case study:

\begin{lstlisting}[basicstyle=\ttfamily\fontsize{6pt}{7pt}\selectfont]
// Pool0.bin of kernel.neff (before), bytes 0-11
slot 67:  43 10 00 00 05 03 13 03 1c 00 00 00   TENSOR_SCALAR_ARITH_OP  wait S[3] >= 28; then S[3] += 1  (join marker)
slot 68:  d4 10 00 00 05 03 00 00 1d 00 00 00   PSEUDO_DMA_DIRECT2D     wait S[3] >= 29                  (reads the output)

// Pool0.bin of final.neff (after)
slot 68:  85 10 00 00 00 00 15 03 09 00 00 00   KERNEL_CALL             operator 0; then S[3] += 9 (SEM_ADD_IMM_COMPLETE)
slot 69:  d4 10 00 00 05 03 00 00 1e 00 00 00   PSEUDO_DMA_DIRECT2D     wait S[3] >= 30
\end{lstlisting}

The main challenge of injector is keeping track of the semaphore. The 9 markers of this call site each added 1 to S[3]. Therefore, the replaced \code{KERNEL\_CALL} instruction needs to add 9 when the operator finishes. Before injection, a wait for a later GPSIMD Engine instruction also implied that the markers had finished. After the renumbering, every wait covers the same instruction as before and also the \code{KERNEL\_CALL} of every earlier call site.

\subsection{Runtime: What Happens Under the Hood}
\label{sec:runtime}

When the Neuron runtime loads the NEFF, it reads \code{gpsimd\_kernels\_def}, and the firmware copies \code{ops.so} from HBM into the local RAM of all 8 cores with the iDMA. At each \code{KERNEL\_CALL}, the firmware calls the operator's entry function on all 8 cores. The entry function reads the core index from \code{PRID} and calls the operator.

The cores reach SBUF and HBM through address-translation windows. \code{gpsimd-run} (a C++ program on libnrt) programs one window over PCIe BAR after it loads the NEFF. During a call, the runtime borrows the iDMA channel from the firmware and restores the channel's descriptor pointer at the end of the call. After every execution, each core writes a receipt: its processor id, the number of calls, the call sites it ran, its cycle count and its status. The receipts prove that the operator ran on the 8 GPSIMD cores instead on the CPU host.

%=====================================================================
\section{Accelerating Engram in DeepSeek-V4.1-Flash with the GPSIMD Engine}
\label{sec:case}

\begin{figure}[!htbp]
  \centering
  \includegraphics[width=0.85\linewidth]{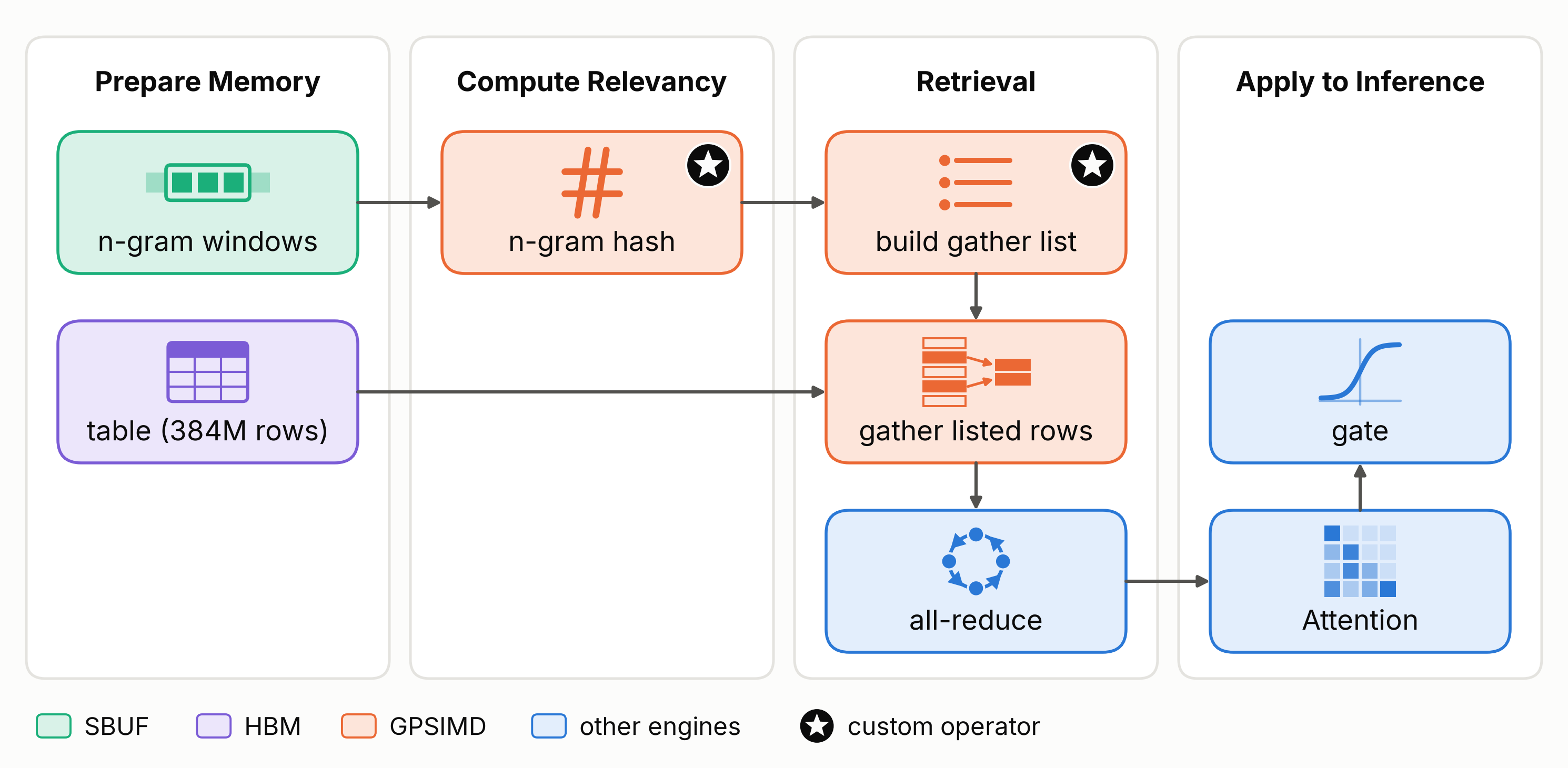}
  \caption{The Engram computation of one layer in the four-step memory processing pipeline of~\citep{he2026memory}, split between the GPSIMD engine and the other engines. The colors show where each step runs or where its data is stored, and $\bigstar$ marks our two custom operators. The gather is NKI code, but the GPSIMD engine issues its indirect DMAs.}
  \label{fig:engram_split}
\end{figure}

In this case study, we accelerate the Engram blocks of the DeepSeek-V4.1-Flash model deployed Trainium 3 in vLLM~\citep{vllm_neuron_ds41} with two custom GPSIMD operators. Figure~\ref{fig:engram_split} shows how one Engram layer is split between the GPSIMD engine and the rest of resources in the Trainium instance. As illustrated in the previous section, we offload the n-gram hashing and the table retrieval to the GPSIMD engine, and the attention and gating network is on the other engines. NKI can already handle indirect DMA gather with GPSIMD engine. Hence, we write custom operators only for n-gram hashing and the gather index list computation.

The baseline is the hand-optimized NKI kernel of Engram in vLLM Neuron. In the DeepSeek V4.1 Flash, the Engram is splitted into 64 ranks (6.0M rows, 3.1\,GB per rank). Each rank only reads the retrieved embeddings that is belong to the shard of the rank, and an all-reduce over the 64 ranks. For a fair comparison, each experiment copies the NKI modules for Engram without changes. For the speed testing, we did not load the real Engram weights. The table shard has the deployed size (6,000,097 rows, 3.1\,GB) with random values, and the row ids follow the hash distribution of the model. The latency is collected using neuron-explorer by measuring the total execution time in the profile.

\subsection{n-gram Hashing}
\label{sec:hash}

The baseline computes the row ids of both Engram layers (48 row ids per token) once per decode step. A decode step of 10 sequences verifies 6 drafted tokens per sequence. Thus, the hash runs on $T = 60$ rows. To compute the n-gram hash, Engram applies the following equations to the compressed ids $x_0, \dots, x_3$ of a token ($x_0$) and its 3 preceding tokens:
\[
  v_n = \bigoplus_{s=0}^{n-1} x_s \cdot m_s, \qquad
  \mathrm{id}_{n,j} = (v_n \bmod p_{n,j}) + o_{n,j}, \qquad n \in \{2, 3, 4\},\ j = 1, \dots, 8,
\]
where $\oplus$ is the bitwise XOR, $m_s$ is an odd multiplier of the layer for token $s$, and $p_{n,j}$ is a large prime number (e.g., 16,000,057). Each pair $(n, j)$ is a \emph{hash column}. The 24 hash columns of a layer share one table of 384M rows, and each hash column owns its own block of $p_{n,j}$ consecutive rows. When computing the hash id, we need to first get the index of the row inside the block of hash column $(n, j)$ by calculating the modulo, then add the offset $o_{n,j}$ (the index of the first row of this block, which is the sum of the sizes of the previous blocks).

The products $x_s \cdot m_s$ and their XOR $v_n$ have up to 63 bits. Hence, the hash requires 64-bit integer arithmetic. However, the Vector and Scalar Engines support only 32-bit integer arithmetic natively. The baseline NKI kernel therefore works on 32-bit pieces. It precomputes a product table, which holds every compressed id multiply with every multiplier of both layers as two 32-bit words. The kernel reads each product from this table with an indirect DMA and then apply bitwise XOR. It then computes $v_n \bmod p_{n,j}$ with Horner's rule, which builds a number from its digits one at a time. The baseline splits the 63-bit $v_n$ into nine 7-bit chunks $c_0, \dots, c_8$, with $c_0$ the most significant, so that
\[
  v_n = \big(\cdots\big((c_0 \cdot 2^7 + c_1) \cdot 2^7 + c_2\big) \cdots\big) \cdot 2^7 + c_8 ,
\]
and takes the remainder after every step:
\[
  r_0 = c_0, \qquad r_k = (r_{k-1} \cdot 2^7 + c_k) \bmod p_{n,j} \quad (k = 1, \dots, 8), \qquad v_n \bmod p_{n,j} = r_8 .
\]
Since $r_{k-1} < p_{n,j} < 2^{24}$, every intermediate value $r_{k-1} \cdot 2^7 + c_k$ is below $2^{31}$ and fits in a 32-bit integer. This is not efficient because each step takes multiple instructions throughout the Horner rule, and they are dependent on each other. With the product lookups and the XORs, the arithmetic part of the hash is 230 instructions in NKI: 166 on the Vector Engine, 58 GPSIMD built-in instructions and 6 DMAs. 

To resolve this issue, we write a custom GPSIMD operator, \code{ngram\_hash}, that computes all 48 row ids of a token with exact 64-bit arithmetic in registers. Each of the 8 cores takes its share of the row embeddings and fetches their n-gram windows, the $p_{n,j}$ and $m_{s}$ into its local data RAM with iDMA. The GPSIMD engine can then do direct 64-bit arithmetics to follow the computations exactly as illustrated in the n-gram hashing formula, instead of having a workaround. The core C++ code looks like the following:

\begin{lstlisting}[language=C++]
// (token id x multiplier) mod 2^64 as two words: MULUH and MULL, then XOR into the rolling value
rh ^= mulhi(cid, ml) + cid * mh;
rl ^= cid * ml;
// v = rh * 2^32 + rl (v < 2^63) mod p, with mu = floor(2^64 / p) from the host:
// Barrett reduction, the quotient estimate is at most 3 too small
const uint64_t mid = (uint64_t)lo * muh + (uint64_t)hi * mul + mulhi(lo, mul);
const uint32_t q = hi * muh + (uint32_t)(mid >> 32);
uint32_t r = lo - q * p;
if (r >= p) r -= p;
if (r >= p) r -= p;
if (r >= p) r -= p;
\end{lstlisting}

\begin{figure}[!htbp]
  \centering
  \includegraphics[width=0.72\linewidth]{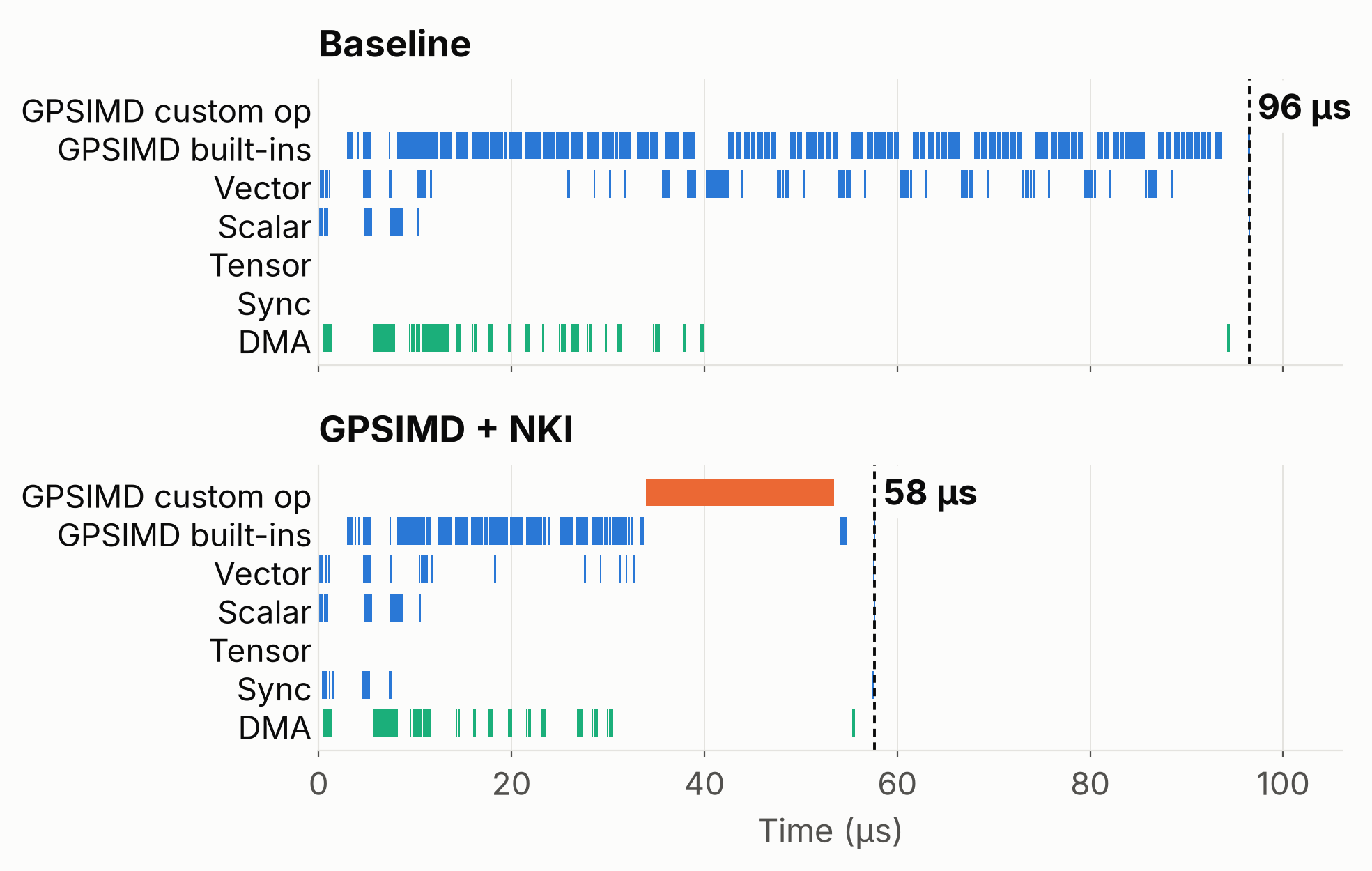}
  \caption{One profiled execution of the hashing at decode (10 sequences, $T = 60$). The n-gram window collection (first 34\us) is the same NKI code in both versions. The arithmetic of the baseline takes the next 62\us; the GPSIMD operator takes 19.5\us.}
  \label{fig:tl_hash}
\end{figure}

Figure~\ref{fig:tl_hash} shows where the time goes. Both versions spend the first 34\us collecting the n-gram windows with the same NKI code. After that, the baseline runs a long chain of small instructions on the Vector Engine and GPSIMD built-in instructions for 62\us, while the GPSIMD version runs one 19.5\us \code{KERNEL\_CALL}. Without counting semaphore waits, the GPSIMD version issues 107 instructions instead of 324. At decode, the hash drops from 86.95\us to 48.45\us (1.79$\times$).

\subsection{Table Retrieval}
\label{sec:retrieval}

After the hashing, each rank reads the rows of the table shard, and an all-reduce over the 64 ranks (Figure~\ref{fig:engram_split}). We studied and experimented on both a prefill chunk of $T = 1024$ and $T = 2048$ tokens, and also on decode. To retrieve the embeddings, rank $r$ computes
\[
  \mathrm{emb}[t, c] =
  \begin{cases}
    E_r[\mathrm{id}_{t,c}] & \text{if } 0 \le \mathrm{id}_{t,c} < R, \\
    0 & \text{otherwise,}
  \end{cases}
\]
where $\mathrm{id}_{t,c}$ is the row id of token $t$ in hash column $c$, $E_r$ is the table shard of rank $r$, and $R$ is its number of rows. We call the lookups whose row id falls inside the shard \emph{hits}.

The number and the positions of the hits are known only at run time. However, NKI needs static shapes: the size of every DMA is fixed when the kernel is compiled. The baseline NKI kernel (\code{engram\_gather}) therefore fetches every row. It computes a mask that marks the hits, issues one indirect DMA of 128 rows for every 128-token tile and hash column, and multiplies the fetched rows by the mask. Consequently, it fetches all rows to keep about 1.5\% of them. This is not efficient because the cost of an indirect DMA grows with its number of indices, not with its number of useful rows, incurring extra unnecessary memory access overhead.

We write the custom operator \code{engram\_hits} to resolve this issue on GPSIMD engine. It builds a \emph{gather list} of the hits, so that DMA fetches only those rows. Each of the 8 cores scans the row ids of its 16 partitions and writes every hit as a pair: the row to read from the shard and the row of \code{emb} to write. An exclusive prefix sum over the per-core counts gives each core its offset. Thus, the pairs of the 8 cores form one contiguous gather list. Core 0 pads the list to a multiple of 128 and writes the number of 128-pair chunks. NKI loads this number into a register and runs a loop with this run-time trip count, with one gather and one scatter per chunk. The core of the operator and the NKI code that calls it are:

\begin{lstlisting}[language=C++]
// GPSIMD: each core scans its partitions (all tiles, all hash columns) after one iDMA fetch
for (uint32_t i = 0; i < NT; ++i)
    for (uint32_t j = 0; j < NCOL; ++j) {
        const uint32_t v = (uint32_t)r[i * NCOL + j];       // row id relative to the shard
        if (v < ROWW) { src[n] = v; dst[n] = (i * 128u + p) * NCOL + j; ++n; }
    }
const uint32_t off = c.exscan_add(n, &total);               // place each core's list
// one iDMA per list; core 0 pads to a multiple of 128 and writes the number of chunks
\end{lstlisting}

\begin{lstlisting}[language=Python,style=py]
# NKI: zero-fill emb, then one gather and one scatter of 128 rows per chunk of hits
gp.call("engram_hits", ins=[ids_sb], outs=[hits_sb, cnt_sb])
reg = nisa.register_alloc(); nisa.register_load(reg, cnt_sb)
nl.fori_loop(0, reg, chunk, step=1)     # chunk(b): load 128 (table row, emb row) pairs, gather, scatter
\end{lstlisting}

\begin{figure}[!htbp]
  \centering
  \includegraphics[width=0.72\linewidth]{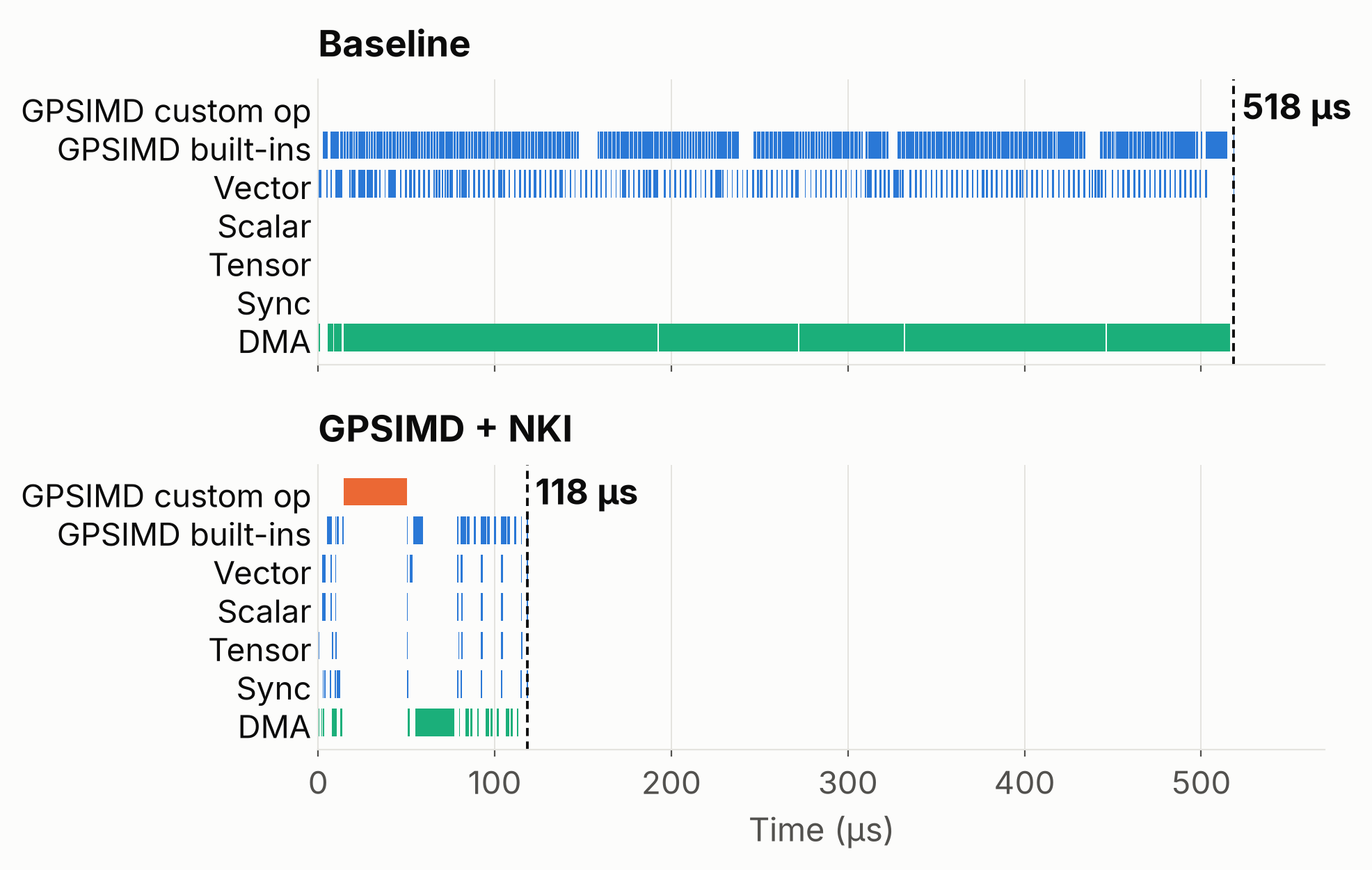}
  \caption{One profiled execution of the table retrieval ($T = 1024$). Top: the masked indirect DMAs of the baseline. Bottom: the \code{engram\_hits} operator (orange), then the gathers and scatters of the hit rows.}
  \label{fig:tl_gather}
\end{figure}

Figure~\ref{fig:tl_gather} shows where the time goes. In the baseline, the GPSIMD Engine is busy with descriptor generation and the Vector Engine with the mask arithmetic for the whole retrieval. In the GPSIMD version, one 36\us operator call replaces this work, and the hits fill only 3 chunks. Therefore, NKI needs 6 indirect DMAs instead of 192. The table retrieval drops from 506.9\us to 102.4\us (4.95$\times$).

\subsection{Results Overview and End-to-End Speedup}
\label{sec:results}

Table~\ref{tab:results} lists all results, including both the per-computation speedups that we discussed in Sections~\ref{sec:hash} and~\ref{sec:retrieval} and the end-to-end speedup of the Engram module. For the whole Engram module (the n-gram hashing and both Engram layers), the two custom operators give a 1.05$\times$ speedup at decode and a 1.14$\times$ speedup on prefill chunks of both 1024 and 2048 tokens.

The end-to-end speedup comes only from the two steps that moved to the GPSIMD cores. In the baseline, the n-gram hashing and the table retrieval take about one fifth of the Engram time on prefill chunks and about one seventh at decode. The attention and the gate take the rest, and they run the same NKI code in both versions. Hence, even if the two steps took no time at all, the Engram module would be at most about 1.27$\times$ faster on prefill chunks and 1.17$\times$ faster at decode. On prefill chunks, the GPSIMD operators remove more than half of the time of the two steps, and the module reaches 1.14$\times$ against this limit of 1.27$\times$.

\begin{figure}[!htbp]
  \centering
  \includegraphics[width=0.72\linewidth]{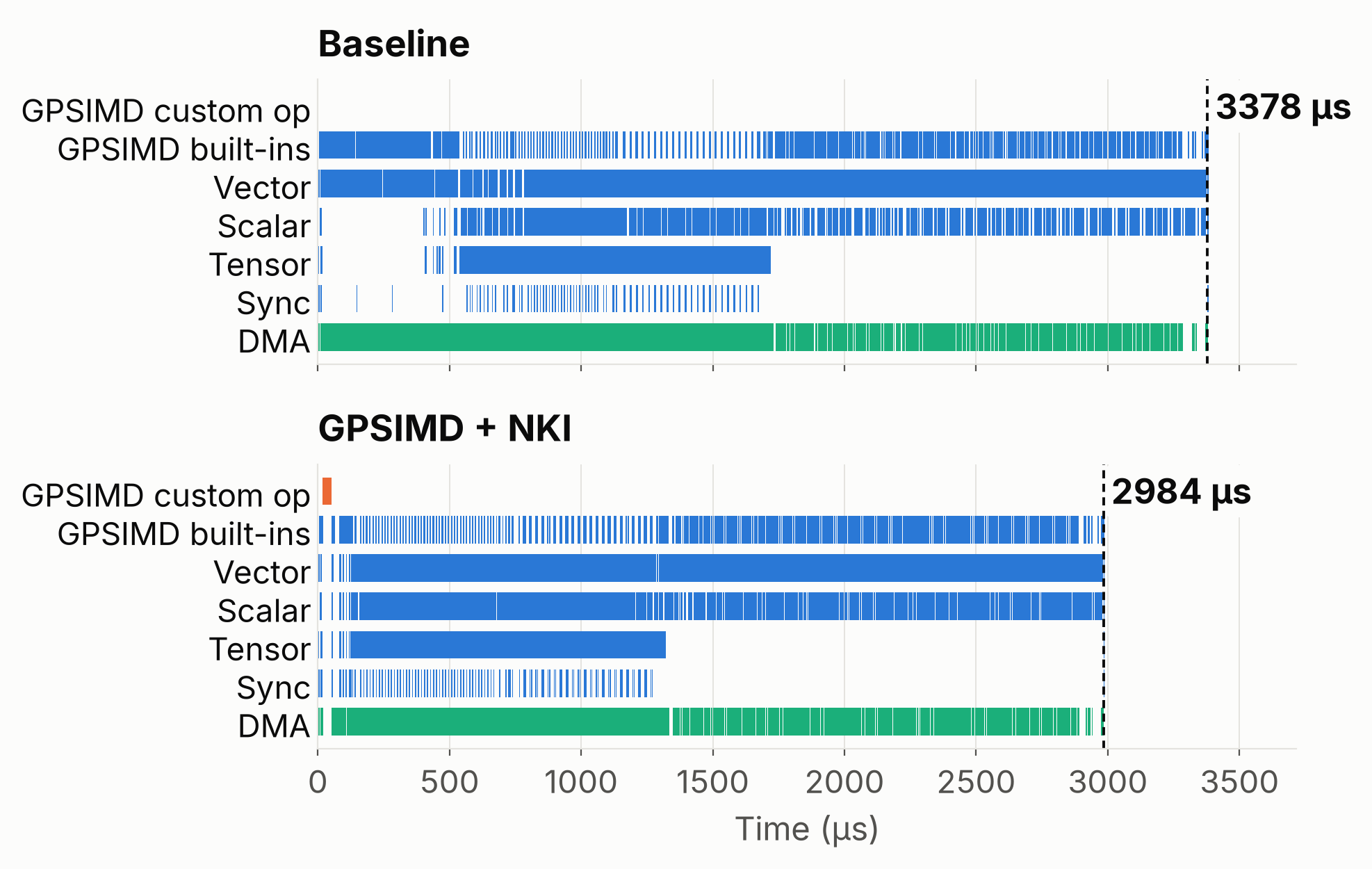}
  \caption{One profiled execution of one Engram layer ($T = 1024$). The GPSIMD operator (orange) shortens the table retrieval. Thus, the attention GEMM starts earlier.}
  \label{fig:tl_readout}
\end{figure}

Most of the time saved on prefill chunks comes from the table retrieval, and Figure~\ref{fig:tl_readout} shows how it reaches the end of the layer. The attention GEMM on the Tensor Engine multiplies the retrieved embeddings by its weights. Therefore, it cannot start until the table retrieval has written all of the retrieved embeddings. With the GPSIMD operator, the table retrieval ends earlier, and the attention GEMM and the gate after it start earlier by the same amount. Consequently, on prefill chunks, the time saved in the table retrieval shortens the whole Engram layer by about the same amount. At decode, the baseline already overlaps the table retrieval with the weight loads of the attention GEMM. Thus, only part of the time saved in the table retrieval shortens the layer (1.02$\times$), and about half of the decode speedup comes from the n-gram hashing.

This speedup is nontrivial since Engram runs for every token of every forward step. More broadly, the result shows that irregular integer steps, which the dense engines handle poorly, can move to the GPSIMD cores inside an existing NKI kernel and give an end-to-end gain without any change to the dense computation around them.

Engram may also take a larger share of the latency in future models. The Engram paper keeps the table in host DRAM: the host computes the row ids and prefetches the rows over PCIe while the GPU computes the preceding blocks~\citep{engram,engram_code}. Our implementation keeps the table in HBM, sharded by row over the 64 ranks, and computes the hash and the retrieval on the device in every forward step. DeepSeek-V4.1-Flash uses Engram in only 2 of its 40 layers, since the Engram paper notes that the placement of the Engram layers must satisfy system latency constraints as well as modeling ones. However, the best models in experiments give about 20 to 25\% of the sparse parameter budget to Engram and the rest to the MoE experts. It also finds that splitting the same table over two Engram layers improves the model slightly over a single layer. If future DeepSeek models follow these findings with more Engram layers, the n-gram hashing and the table retrieval will take a larger share of the latency.

\begin{table}[!htbp]
  \centering
  \caption{Latency of the Engram computations on one NeuronCore, for the baseline and with GPSIMD custom operators (GPSIMD + NKI).}
  \label{tab:results}
  \small
  \begin{tabular}{@{}P{3.3cm}lrrr@{}}
    \toprule
    \textbf{Engram computation} & \textbf{Config} & \textbf{Baseline} & \textbf{GPSIMD + NKI} & \textbf{Speedup} \\
    \midrule
    n-gram hashing & decode, 10 sequences ($T = 60$) & 86.95\us & 48.45\us & \textbf{1.79$\times$} \\
     & prefill chunk, $T = 1024$ & 531.1\us & 449.7\us & 1.18$\times$ \\
     & prefill chunk, $T = 2048$ & 1055.6\us & 890.6\us & 1.19$\times$ \\
    \midrule
    table retrieval & decode ($T = 60$) & 62.1\us & 24.7\us & \textbf{2.51$\times$} \\
     & prefill chunk, $T = 1024$ & 506.9\us & 102.4\us & \textbf{4.95$\times$} \\
     & prefill chunk, $T = 2048$ & 980.5\us & 200.7\us & \textbf{4.89$\times$} \\
    \midrule
    one Engram layer & decode ($T = 60$) & 697.1\us & 681.0\us & 1.02$\times$ \\
     & prefill chunk, $T = 1024$ & 3362.9\us & 2968.9\us & \textbf{1.13$\times$} \\
     & prefill chunk, $T = 2048$ & 6766.4\us & 5926.4\us & \textbf{1.14$\times$} \\
    \midrule
    whole Engram (hashing + 2 layers) & decode ($T = 60$) & 1481.2\us & 1410.5\us & \textbf{1.05$\times$} \\
     & prefill chunk, $T = 1024$ & 7256.9\us & 6387.5\us & \textbf{1.14$\times$} \\
     & prefill chunk, $T = 2048$ & 14588.4\us & 12743.4\us & \textbf{1.14$\times$} \\
    \bottomrule
  \end{tabular}
\end{table}

%=====================================================================
\section{What Other Operations Can GPSIMD Help With?}
\label{sec:other}

We tried the GPSIMD custom kernel in other scenarios as well. Although some kernels may not be the bottleneck of inference or unrelated to ML workloads at this moment, we believe that this can broaden our scope of future algorithm design instead of sticking to the dense operations that Tensor/Vector/Scalar engines are good at. Any step that is integer-heavy, data-dependent, irregular, and small in data compared with the dense work around it is a candidate (see Section~\ref{sec:trainium}). Before Engram, we used the same toolchain on kernels of the NKI Library~\citep{nki_library}. Against the library's NKI implementations, we measured the speedups in Table~\ref{tab:other}.

\begin{table}[h]
  \centering
  \caption{Speedups of GPSIMD operators over the NKI Library implementations.}
  \label{tab:other}
  \small
  \begin{tabular}{lr}
    \toprule
    \textbf{Kernel} & \textbf{Speedup over optimized NKI} \\
    \midrule
    argsort & 6.6$\times$ to 40$\times$ \\
    BM25 scoring for RAG & 15$\times$ to 27$\times$ \\
    scatter-add with duplicate indices & 4.45$\times$ \\
    \bottomrule
  \end{tabular}
\end{table}

%=====================================================================
\section{Limitations}
\label{sec:limitations}

Currently, our toolchain supports only part of what the GPSIMD cores can do. The main gaps are listed below.

\paragraph{No compiler support for vector instructions.} Each GPSIMD core has a 512-bit vector unit, and the unit is enabled while our operators run. However, the only Xtensa configuration in the Neuron SDK (\code{ncore2gp}) describes the Trainium~1 core. Trainium~3 encodes the vector instructions differently. Consequently, vector code generated by \code{xt-clang} hangs the core. The toolchain therefore compiles scalar base-ISA code only, and \code{gpsimd-cc} rejects any library that contains vector instructions.

\paragraph{No hardware floating point.} The floating-point instructions of \code{ncore2gp} also hang the core, and we have not found their Trainium~3 encodings. Floating point therefore runs in software (\code{gp::f32}): a float32 addition takes 117 cycles and a multiplication 97 cycles. Steps dominated by floating-point math still belong on the Vector and Scalar Engines.

\paragraph{Tile addresses fixed at build time.} We did not find a way for the operator to read the \code{KERNEL\_CALL} operands at run time. Therefore, the SBUF addresses of the tiles are compiled into \code{ops.so}. The library must be rebuilt whenever the kernel is recompiled, and it must fit in 64\,KB of code and about 7\,KB of static data. 

An official support on GPSIMD custom kernel should resolve these issues.

%=====================================================================
\section*{Acknowledgments}

This work is supported by the AWS Build on Trainium program at UCLA. We used AutoHelix~\citep{autohelix} as the agentic loop harness to optimize the kernel, with special modifications to add constraints based on human experts. We would like to thank Ellen Hermansen from the Amazon Research Initiative team for providing AWS access and promotional credits, Amit Kachroo for providing technical insights in the Amazon Science Hub program, and James Hamilton for the continuous support on providing latest Trainium instances. 

\bibliographystyle{plain}
\bibliography{references}

@article{he2026memory,
  author  = {Zifan He and Rui Ma and Yizhou Sun and Jason Cong},
  title   = {Understand and Accelerate Memory Processing Pipeline for Large Language Model Inference},
  journal = {arXiv preprint arXiv:2603.29002},
  year    = {2026},
  url     = {https://arxiv.org/abs/2603.29002}
}

@misc{heterollm,
  author       = {Zifan He},
  title        = {{HeteroLLM}: {GPU-FPGA} Heterogeneous Inference Framework},
  year         = {2026},
  howpublished = {\url{https://github.com/OswaldHe/HeteroLLM}},
  note         = {Engram implementation in \texttt{kernels/deepseek\_engram}}
}

@misc{deepseek_v41_flash,
  author       = {{DeepSeek-AI}},
  title        = {{DeepSeek-V4.1-Flash}},
  year         = {2026},
  howpublished = {\url{https://huggingface.co/deepseek-ai/DeepSeek-V4.1-Flash}},
  note         = {Model card on Hugging Face}
}

@article{engram,
  author  = {{DeepSeek-AI}},
  title   = {Conditional Memory via Scalable Lookup: A New Axis of Sparsity for Large Language Models},
  journal = {arXiv preprint arXiv:2601.07372},
  year    = {2026},
  url     = {https://arxiv.org/abs/2601.07372}
}

@misc{engram_code,
  author       = {{DeepSeek-AI}},
  title        = {{Engram} Reference Implementation},
  year         = {2026},
  howpublished = {\url{https://github.com/deepseek-ai/Engram}},
  note         = {\texttt{engram\_demo\_v1.py}}
}

@misc{neuron_trn3_arch,
  author       = {{AWS Neuron}},
  title        = {{Trainium3} Architecture},
  year         = {2026},
  howpublished = {\url{https://awsdocs-neuron.readthedocs-hosted.com/en/latest/general/arch/neuron-hardware/trainium3.html}},
  note         = {AWS Neuron documentation}
}

@misc{neuron_trn3_nki_arch,
  author       = {{AWS Neuron}},
  title        = {{Trainium3} Architecture Guide for {NKI}},
  year         = {2026},
  howpublished = {\url{https://awsdocs-neuron.readthedocs-hosted.com/en/latest/nki/guides/architecture/trainium3_arch.html}},
  note         = {AWS Neuron documentation}
}

@misc{nki_about,
  author       = {{AWS Neuron}},
  title        = {About {NKI}},
  year         = {2026},
  howpublished = {\url{https://awsdocs-neuron.readthedocs-hosted.com/en/latest/nki/get-started/about/index.html}},
  note         = {AWS Neuron documentation}
}

@misc{vllm_neuron_ds41,
  author       = {Zifan He},
  title        = {{DeepSeek-V4.1-Flash} Port for {Trainium 3} in vllm-neuron},
  year         = {2026},
  howpublished = {\url{https://github.com/OswaldHe/vllm-neuron/tree/deepseek_v41_trn3/contrib/models/deepseek_v41/src}},
  note         = {Branch \texttt{deepseek\_v41\_trn3}}
}

@misc{neff_doc,
  author       = {Zifan He and Jason Cong},
  title        = {Accelerating {AWS Trainium} Below Kernel Level: {NEFF}-based Performance Optimization},
  year         = {2026},
  howpublished = {\url{https://github.com/OswaldHe/neuron-neff-doc}}
}

@misc{nki_library,
  author       = {{AWS Neuron}},
  title        = {{NKI Library}},
  year         = {2026},
  howpublished = {\url{https://github.com/aws-neuron/nki-library}}
}

@misc{autohelix,
  author       = {{AWS Labs}},
  title        = {{AutoHelix}},
  year         = {2026},
  howpublished = {\url{https://github.com/awslabs/AutoHelix}}
}

@misc{neuron_customops,
  author       = {{AWS Neuron}},
  title        = {Neuron Custom {C++} Operators},
  year         = {2026},
  howpublished = {\url{https://awsdocs-neuron.readthedocs-hosted.com/en/v2.33.0/archive/neuron-customops/index.html}},
  note         = {AWS Neuron documentation, Neuron SDK 2.33 (archived)}
}

\end{document}